\documentclass[a4paper]{article}

\usepackage{INTERSPEECH2021}

\usepackage{hyperref,multirow,url}

\newcommand{\scell}[2][c]{%
  \begin{tabular}[#1]{@{}c@{}}#2\end{tabular}}
  
\title{KazakhTTS: An Open-Source Kazakh Text-to-Speech Synthesis Dataset}
\name{Saida Mussakhojayeva, Aigerim Janaliyeva, Almas Mirzakhmetov,\\Yerbolat Khassanov and Huseyin Atakan Varol}
\address{
  Institute of Smart Systems and Artificial Intelligence (ISSAI),\\Nazarbayev University, Nur-Sultan, Kazakhstan}
\email{\{saida.mussakhojayeva, aigerim.janaliyeva, almas.mirzakhmetov,\\ yerbolat.khassanov, ahvarol\}@nu.edu.com}

\begin{document}

\maketitle
\begin{abstract}
This paper introduces a high-quality open-source speech synthesis dataset for Kazakh, a low-resource language spoken by over 13 million people worldwide.
The dataset consists of about 93 hours of transcribed audio recordings spoken by two professional speakers (female and male).
It is the first publicly available large-scale dataset developed to promote Kazakh text-to-speech (TTS) applications in both academia and industry.
In this paper, we share our experience by describing the dataset development procedures and faced challenges, and discuss important future directions.
To demonstrate the reliability of our dataset, we built baseline end-to-end TTS models and evaluated them using the subjective mean opinion score (MOS) measure.
Evaluation results show that the best TTS models trained on our dataset achieve MOS above 4 for both speakers, which makes them applicable for practical use.
The dataset, training recipe, and pretrained TTS models are freely available\footnote{\label{ft:github}\url{https://github.com/IS2AI/Kazakh_TTS}}.
\end{abstract}

\section{Introduction}
Text-to-speech (TTS) systems are essential in many applications, such as navigation, announcement, smart assistants, and other speech-enabled devices.
For any language, it is also necessary to ensure accessibility for the visually-impaired, and availability of human-machine interaction without requiring visual and tactile interfaces~\cite{taylor2009text}. To build a robust TTS system, a sufficiently large and high-quality speech dataset is required.
In order to address this, we developed a large-scale open-source speech dataset for the Kazakh language.
We named our dataset KazakhTTS, and it is primarily geared to build TTS systems.

Kazakh is the official language of Kazakhstan, and it is spoken by over 13 million people worldwide\footnote{\url{https://www.ethnologue.com/language/kaz}}, including countries, such as China and Russia.
It is an agglutinative language with vowel harmony belonging to the family of Turkic languages.
Kazakh is a low-resource language and is considered to be endangered due to multiple factors, primarily the dominance of the Russian and English languages in education and administration~\cite{dave2007kazakhstan}.
Therefore, there is a growing awareness of the importance of increasing the number of Kazakh speakers reflected in many language rescue initiatives launched by the government.

Currently, there is no Kazakh speech dataset of sufficient size and quality for building TTS systems, especially recently proposed end-to-end (E2E) neural architectures~\cite{DBLP:journals/corr/WangSSWWJYXCBLA17,DBLP:conf/iclr/SoteloMKSKCB17,DBLP:conf/icml/ArikCCDGKLMNRSS17,DBLP:conf/icml/Skerry-RyanBXWS18,DBLP:conf/icassp/ShenPWSJYCZWRSA18,DBLP:conf/nips/RenRTQZZL19,DBLP:journals/corr/abs-1809-08895}.
This work aims to fill this gap by introducing the KazakhTTS dataset.
To the best of our knowledge, it is the first large-scale open-source dataset developed for building Kazakh TTS systems.
Our dataset contains around 93 hours of high-quality speech data read by two professional speakers (36 hours by female and 57 hours by male).
The dataset was carefully annotated by native transcribers and covers most of the daily-use Kazakh words.
The KazakhTTS is freely available for both academic and commercial use under the Creative Commons Attribution 4.0 International License\footnote{\url{https://creativecommons.org/licenses/by/4.0/}}.

With the help of KazakhTTS, we plan to promote Kazakh language use in speech-based digital technologies and to advance research in Kazakh speech processing.
We believe that our dataset will be a valuable resource for the TTS research community, and our experience will benefit other researchers planning to develop speech datasets for low-resource languages. 
Although the primary application domain of KazakhTTS is speech synthesis, it can also be used to aid other related applications, such as automatic speech recognition (ASR) and speech-to-speech translation.

To demonstrate the reliability of KazakhTTS, we built two baseline Kazakh E2E-TTS systems based on Tacotron 2~\cite{DBLP:conf/icassp/ShenPWSJYCZWRSA18} and Transformer~\cite{DBLP:journals/corr/abs-1809-08895} architectures for each speaker.
We evaluated these systems using the subjective mean opinion score (MOS) measure.
The experiment results showed that the best TTS models built using our dataset achieve 4.5 and 4.1 in MOS for the female and male speakers, respectively, which assures the usability for practical applications.
In addition, we performed a manual analysis of synthesized sentences to identify the most frequent error types.
The dataset, reproducible recipe, and pretrained models are publicly available\textsuperscript{\ref{ft:github}}.

The rest of the paper is organized as follows.
Section 2 briefly reviews related works for TTS dataset construction.
Section 3 describes the KazakhTTS construction procedures and reports the dataset specifications.
Section 4 explains the TTS experimental setup and discusses obtained results.
Section 5 concludes the paper and highlights future research directions.


\section{Related works}
The recent surge of speech-enabled applications, such as virtual assistants for smart devices, has attracted substantial attention from both academia and industry to the TTS research~\cite{DBLP:journals/corr/WangSSWWJYXCBLA17,DBLP:conf/iclr/SoteloMKSKCB17,DBLP:conf/icml/ArikCCDGKLMNRSS17}.
Consequently, many large-scale datasets have been collected~\cite{ljspeech17,DBLP:conf/interspeech/ZenDCZWJCW19,DBLP:journals/corr/abs-2010-11567,veaux2016superseded}, and challenges have been organized to systematically compare different TTS technologies~\cite{DBLP:conf/interspeech/BlackT05}.
However, these datasets and competitions are mostly restricted to resource-rich languages, such as English, Mandarin, and so on.

To create a speech corpus suitable for developing TTS systems in low-resource settings, methods based on unsupervised~\cite{DBLP:conf/icml/RenTQZZL19}, semi-supervised~\cite{DBLP:conf/icassp/ChungWHZS19}, and cross-lingual transfer learning~\cite{DBLP:conf/interspeech/ChenTYL19} have been developed.
In these approaches, raw audio files are automatically annotated by using other systems, such as ASR, or available resources from other languages are utilized, especially from phonologically close languages.
Although TTS systems produced using these approaches have achieved good results, their quality is usually insufficient for practical applications.
Furthermore, these approaches require some in-domain data or other systems, such as ASR, speech segmentation, and speaker diarisation, which might be unavailable for low-resource languages.

An alternative approach is to record and manually annotate audio recordings.
This is considered costly, since cumbersome manual work is required.
Nevertheless, the produced dataset will be more reliable.
Recently, several Kazakh speech corpora have been developed to accelerate speech processing research in this language.
For example, the first open-source Kazakh speech corpus (KSC) containing over 332 hours of transcribed audio recordings was presented in~\cite{khassanov2020crowdsourced}.
However, the KSC was mainly constructed for ASR applications, and thus, crowdsourced from different speakers, with various background noises and speech disfluencies kept to make it similar to the real-world scenarios.
As such, the quality of audio recordings in the KSC is insufficient for building robust TTS models.
Additionally, in the KSC, the size of recordings contributed by a single speaker is small.
Similarly, the other existing Kazakh speech datasets are either unsuitable or publicly unavailable~\cite{DBLP:conf/emnlp/MakhambetovMYMSS13,DBLP:conf/aciids/MamyrbayevAZTG20}.

\section{Dataset construction}
The KazakhTTS project was conducted with the approval of the Institutional Research Ethics Committee of Nazarbayev University.
Each speaker participated voluntarily and was informed of the data collection and use protocols through a consent form.

\subsection{Text collection and narration}
We started the dataset construction process with textual data collection.
In particular, we manually extracted articles in chronological order from news websites to diversify the topic coverage (e.g., politics, business, sports, entertainment and so on) and eliminate defects peculiar to web crawlers.
The extracted articles were manually filtered to exclude inappropriate content (e.g., sensitive political issues, user privacy, and violence) and stored in DOC format convenient for professional speakers (i.e., font size, line spacing, and typeface were adjustable to the preferences of the speakers). 
In total, we collected over 2,000 articles of different lengths.

To narrate the collected articles, we first conducted an audition among multiple candidates from which one female and one male professional speaker were chosen.
The speaker details, including age, working experience as a narrator, and recording device information, are provided in Table~\ref{tab:spk}.
Due to the COVID-19 pandemic, we could not invite the speakers to our laboratory for data collection.
Therefore, the speakers were allowed to record audio at their own studio at home or office.
The speakers were instructed to read texts in a quiet indoor environment at their own natural pace and style.
Additionally, they were asked to follow orthoepic rules, pause on commas, and use appropriate intonations for sentences ending with a question mark. 
The female speaker recorded audios at her office studio, the recordings were sampled at 44.1 kHz and stored using 16 bit/sample.
The male speaker recorded audios at home studio, the recordings were sampled at 48 kHz and stored using 24 bit/sample.
In total, female and male speakers read around 1,000 and 1,250 articles respectively, out of which around 300 articles are overlapping. 

\begin{table}[t]
  \renewcommand\arraystretch{1.0}
  \setlength{\tabcolsep}{1.2mm}
  \caption{The KazakhTTS speaker information}
  \label{tab:spk}
  \centering
  \begin{tabular}{c|c|c|c|c }
    \toprule
    \textbf{Speaker ID} & \textbf{Gender}   & \textbf{Age}  & \textbf{\scell{Work\\experience}} & \textbf{\scell{Recording\\device}} \\
    \midrule
    F1                  & Female            & 44            & 14 years                          & AKG P120 \\ 
    M1                  & Male              & 46            & 12 years                          & Tascam DR-40 \\
    \bottomrule
  \end{tabular}
\end{table}

\subsection{Segmentation and audio-to-text alignment}
We hired five native Kazakh transcribers to segment the recorded audio files into sentences and align them with text using the Praat toolkit~\cite{boersma2001praat}.
The texts were represented using the Cyrillic alphabet consisting of 42 letters\footnote{Note that the Kazakh writing system differs depending on the region where it is used. In Kazakhstan, the Cyrillic alphabet is used.}.
In addition to letters, the transcripts also contain punctuation marks, such as period (`.'), comma (`,'), hyphen (`-'), question mark (`?'), exclamation mark (`!'), and so on. 
The transcribers were instructed to remove segments with incorrect pronunciation and background noise, trim long pauses at the beginning and end of the segments, and convert numbers and special characters (e.g., `\%', `\$', `+', and so on) into the written forms.
To ensure uniform quality of work among the transcribers, we assigned a linguist to randomly check the tasks completed by transcribers and organize regular ``go through errors" sessions.

To guarantee high quality, we inspected the segmented sentences using our ASR system trained on the KSC dataset~\cite{khassanov2020crowdsourced}.
Specifically, we used the ASR system to transcribe the segments.
The recognized transcripts were then compared with the corresponding manually annotated transcripts.
The segments with a high character error rate (CER) were regarded as incorrectly transcribed, and thus, rechecked by the linguist. 
Lastly, we filtered out all segments containing international words written using a non-Cyrillic alphabet, because speakers usually don't know how to correctly pronounce such words.

\subsection{Dataset specifications}
\begin{figure}[th]
  \centering
  \includegraphics[width=1.0\linewidth,trim={0.575cm 3.625cm 11.125cm 0.275cm},clip=true]{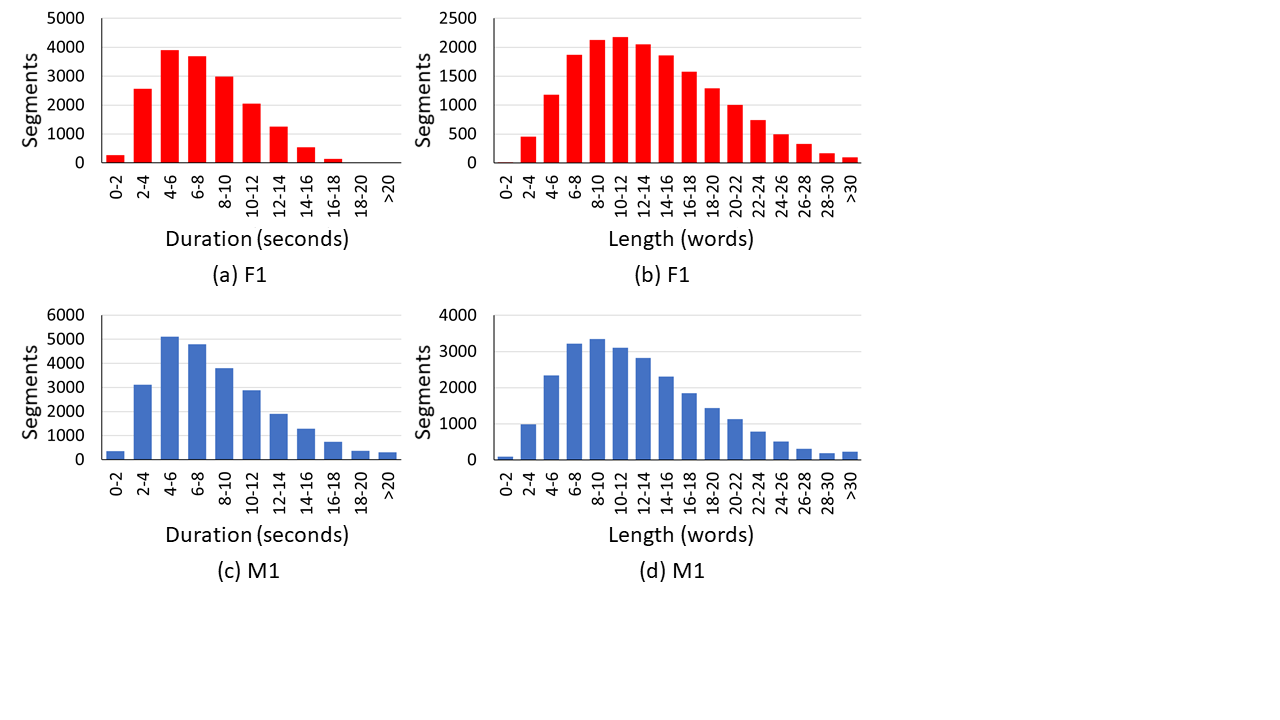}
  \caption{Segment duration and length distributions for female (a and b) and male (c and d) speakers in the KazakhTTS.}
  \label{fig:hist}
\end{figure}

The overall statistics of the constructed dataset are provided in Table~\ref{tab:data}.
In total, the dataset contains around 93 hours of audio consisting of over 42,000 segments.
The distribution of segment lengths and durations are shown in Figure~\ref{fig:hist}.
The whole dataset creation process took around five months, and the uncompressed dataset size is around 15 GB.

The KazakhTTS dataset is organized as follows.
The resources of two professional speakers are stored in two separate folders.
Each folder contains a single metadata file and two subfolders containing audio recordings and their transcripts.
The audio and corresponding transcript filenames are the same, except that the audio recordings are stored as WAV files, whereas the transcripts are stored as TXT files using the UTF-8 encoding.
The naming convention for both WAV and TXT files are as follows \textit{source\_articleID\_segmentID}.
The audio recordings of both speakers have been downsampled to 22.05 kHz, with samples stored as 16-bit signed integers.
The metadata contains speaker information, such as age, gender, working experience, and recording device.

\begin{table}[th]
  \renewcommand\arraystretch{1.1}
  \setlength{\tabcolsep}{5.0mm}
  \caption{The KazakhTTS dataset specifications}
  \label{tab:data}
  \centering
  \begin{tabular}{ l|l|cc }
    \toprule
    \multicolumn{2}{l|}{\multirow{2}{*}{\textbf{Category}}}                 & \multicolumn{2}{c}{\textbf{Speaker ID}}  \\\cline{3-4}
    \multicolumn{2}{l|}{}                                                   & \textbf{F1}   & \textbf{M1} \\
    \midrule
    \multicolumn{2}{l|}{\# Segments}                                        & 17,426        & 24,656   \\\hline
    \multirow{4}{*}{\scell{Segment\\duration}}& Total                       & 36.1 hr       & 57.1 hr \\
                                                                & Mean      & 7.5 sec       & 8.3 sec \\
                                                                & Min       & 1.0 sec       & 0.8 sec \\
                                                                & Max       & 24.2 sec      & 55.9 sec \\\hline
    \multirow{4}{*}{Words}                                      & Total     & 245.4k        & 320.2k \\
                                                                & Mean      & 14.0          & 12.9   \\
                                                                & Min       & 2             & 1 \\
                                                                & Max       & 42            & 75 \\\hline
    \multicolumn{2}{l|}{\# Unique words}                                    & 34.2k         & 42.4k \\
    \bottomrule
  \end{tabular}
\end{table}

\section{TTS experiments}
To demonstrate the utility of the constructed dataset, we built the first Kazakh E2E-TTS systems and evaluated them using the subjective MOS measure.

\subsection{Experimental setup}
We used ESPnet-TTS toolkit~\cite{hayashi2020espnet} to build the E2E-TTS models based on Tacotron 2~\cite{DBLP:conf/icassp/ShenPWSJYCZWRSA18} and Transformer~\cite{DBLP:journals/corr/abs-1809-08895} architectures.
Specifically, we followed the LJ Speech~\cite{ljspeech17} recipe and used the latest ESPnet-TTS developments to configure our model building recipe.
The input for each model is a sequence of characters\footnote{Due to the strong correspondence between word spelling and pronunciation in Kazakh, we did not convert graphemes to phonemes.} consisting of 42 letters and 5 symbols (`.', `,', `-', `?', `!'), and the output is a sequence of acoustic features (80 dimensional log Mel-filter bank features).
To transform these acoustic features into the time-domain waveform samples, we tried different approaches, such as Griffin-Lim algorithm~\cite{DBLP:conf/waspaa/PerraudinBS13}, WaveNet~\cite{DBLP:conf/ssw/OordDZSVGKSK16}, and WaveGAN~\cite{DBLP:conf/icassp/YamamotoSK20} vocoders. 
We found WaveGAN to perform best in our case, and it was used in our final E2E-TTS systems.
We did not apply any additional speech preprocessing, such as filtering, normalization, and so on.

In the Tacotron 2 system, the encoder module was modeled as a single bi-directional LSTM layer with 512 units (256 units in each direction), and the decoder module was modelled as a stack of two unidirectional LSTM layers with 1,024 units.
The parameters were optimized using the Adam algorithm~\cite{DBLP:journals/corr/KingmaB14} with an initial learning rate of $10^{-3}$ for 200 epochs.
To regularize parameters, we set the dropout rate to 0.5.

The Transformer system was modeled using six encoder and six decoder blocks.
The number of heads in the self-attention layer was set to 8 with 512-dimension hidden states, and the feed-forward network dimensions were set to 1,024. 
The model parameters were optimized using the Adam algorithm with an initial learning rate of 1.0 for 200 epochs.
The dropout rate was set to 0.1.

For each speaker, we trained separate E2E-TTS models (i.e., single speaker model).
All models were trained using the Tesla V100 GPUs running on an NVIDIA DGX-2 server.
More details on model specifications and training procedures are provided in our GitHub repository\textsuperscript{\ref{ft:github}}.

\subsection{Experimental evaluation}
To assess the quality of the synthesised recordings, we conducted subjective evaluation using the MOS measure.
We performed a separate evaluation session for each speaker.
In each session, the following three systems were compared: 1) Ground truth (i.e., natural speech), 2) Tacotron 2, and 3) Transformer. 

As an evaluation set, we randomly selected 50 sentences of various lengths from each speaker.
These sentences were not used to train the models.
The selected sentences were manually checked to ensure that each of them is a single complete sentence, and the speaker read them well (i.e., without disfluencies, mispronunciations, or background noise).
The number of listeners was 50 in both evaluation sessions\footnote{In fact, the number of listeners was higher than 50, however, we excluded the ones who did not complete the session till the end and the suspicious listeners who, for example, rated all recordings as excellent, rated all ground truth recordings as bad, and so on.}. 
The listeners were instructed to assess the overall quality, use headphones, and sit in a quiet environment\footnote{Due to the crowdsourced nature of the evaluation process, we can not guarantee that all listeners used headphones and sat in a quiet environment.}.

\begin{figure}[t]
  \centering
  \frame{\includegraphics[width=0.9\linewidth,trim={0.175cm 9.975cm 20.8cm 0.1cm},clip=true]{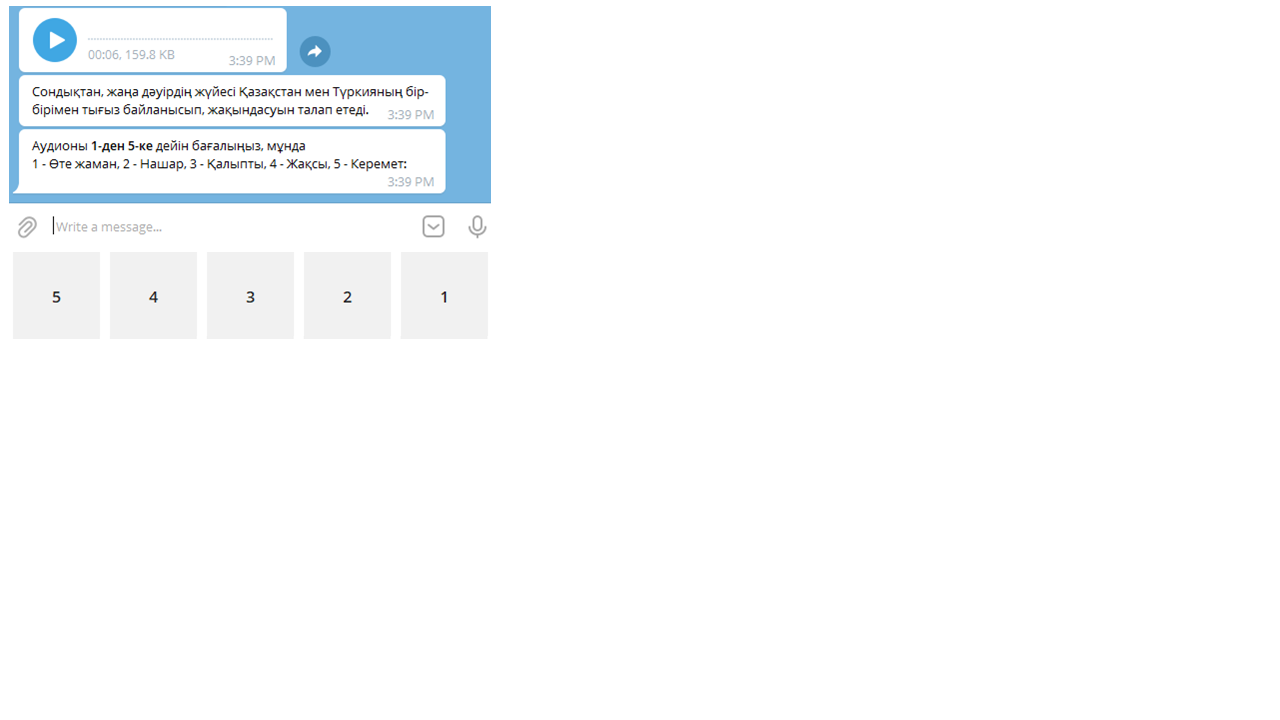}}
  \caption{An example of interaction with the Telegram bot during an evaluation session.}
  \label{fig:tele_bot}
\end{figure}

The evaluation sessions were conducted through the instant messaging platform Telegram~\cite{telegram}, since it is difficult to find native Kazakh listeners on other well-known platforms, such as Amazon Mechanical Turk~\cite{MTurk}.
We developed two separate Telegram bots for each speaker.
The bots first presented a welcome message with the instruction and then started the evaluation process.
During the evaluation, the bots sent a sentence recording with the transcript to a listener and received the corresponding evaluation score (see Figure~\ref{fig:tele_bot}). 
The recordings were rated using a 5-point Likert scale: 5 for excellent, 4 for good, 3 for fair, 2 for poor, and 1 for bad.
Note that in Telegram, to send audio recordings, we had to convert them into MP3 format.

We attracted listeners to participate in the evaluation sessions by advertising the project in social media, news, and open messaging communities on WhatsApp and Telegram.
The listeners were allowed to listen to recordings several times, but they were not allowed to alter previous ratings once submitted.
Additionally, the Telegram bots were keeping track of the listeners' status and ID.
As a result, the listeners could take a break to continue at a later time, and were prevented from participating in the evaluation session more than once.

For all the listeners, the evaluation recordings were presented in the same order and one at a time.
However, at each time step, the bots randomly decided from which system to pick a recording.
As a result, each listener heard each recording once only, and all the systems were exposed to all the listeners.
Each recording was rated at least 8 and 10 times for the female and male speakers, respectively.
At the end of evaluation, the bots thanked the listeners and invited them to fill in an optional questionnaire asking about their age, region (where a listener grew up and learned the Kazakh language), and gender information.
The questionnaire results showed that the listeners varied in gender and region, but not in age (most of them were under 20).

The subjective evaluation results are given in Table~\ref{tab:result}.
According to the results, the best performance is achieved by the Ground truth, as expected, followed by the Tacotron 2, and then the Transformer system for both speakers.
Importantly, the best performing models for both speakers achieved above 4 in the MOS measure and are not too far from the Ground truth, i.e., 4\% and 5\% relative reduction for the female and male speakers, respectively.
These results demonstrate the utility of our KazakhTTS dataset for TTS applications. 

We did not carry out an intensive hyper-parameter tuning for our E2E-TTS models, since it is outside the scope of this work, therefore, we speculate that the quality of models can be further improved.
For example, our model training recipes are based on the LJ Speech, which is tuned for a dataset containing around 24 hours of audio, whereas our speakers' audio sizes are larger.
We leave the exploration of the optimal hyper-parameter settings and detailed comparison of different TTS architectures for the Kazakh language as a future work.

\begin{table}[t]
  \renewcommand\arraystretch{1.1}
  \setlength{\tabcolsep}{4.4mm}
  \caption{Mean opinion score (MOS) results with 95\% confidence intervals}
  \label{tab:result}
  \centering
  \begin{tabular}{ l|cc }
    \toprule
    \multirow{2}{*}{\textbf{System}}    & \multicolumn{2}{c}{\textbf{Speaker ID}} \\\cline{2-3}
                                        & \textbf{F1}                   & \textbf{M1} \\
    \midrule
    Ground truth                        & \textbf{4.726 $\pm$ 0.037}    & \textbf{4.360 $\pm$ 0.050} \\
    Tacotron 2                           & 4.535 $\pm$ 0.049             & 4.144 $\pm$ 0.063 \\
    Transformer                         & 4.436 $\pm$ 0.057             & 3.907 $\pm$ 0.076 \\
    \bottomrule
  \end{tabular}
\end{table}

\subsubsection{Manual analysis}
To identify error types made by E2E-TTS system, we manually analysed a 50-sentence evaluation set for both speakers.
This analysis was conducted only for the Tacotron 2 system which achieved better MOS score than the Transformer.
Specifically, we counted sentences with various error types.
The analysis results are provided in Table~\ref{tab:error}, showing that the most frequent error types for both speakers are mispronunciation and incomplete words.
The mispronunciation errors are mostly due to incorrect stress, and the incomplete word errors mostly occur at the last word of a sentence, where the last letters of the word are trimmed.
Interestingly, the total number of errors in the male speaker's recordings is considerably higher than the total number of errors in the female speaker's recordings.
This might be one of the reasons for the lower MOS score achieved by the male speaker.
This analysis indicates that there is still room for improvement and future work should focus on eliminating these errors.

\begin{table}[t]
  \renewcommand\arraystretch{1.1}
  \caption{Manual analysis of error types made by Tacotron 2}
  \label{tab:error}
  \centering
  \begin{tabular}{ l|cc }
    \toprule
    \multirow{2}{*}{\textbf{Error types}}       & \multicolumn{2}{l}{\textbf{Speaker ID}} \\\cline{2-3}
                                                & \textbf{F1}   & \textbf{M1} \\
    \midrule
    Sentences containing repeated words         & 0             & 0 \\
    Sentences containing skipped words          & 0             & 2 \\
    Sentences containing mispronounced words    & 4             & 8 \\
    Sentences containing incomplete words       & 3             & 9 \\
    Sentences containing long pauses            & 0             & 2 \\
    Sentences containing nonverbal sounds       & 0             & 0 \\\hline
    \textbf{Total}                              & 7             & 21 \\
  \bottomrule
  \end{tabular}
\end{table}

\section{Conclusion}
This paper introduced the first open-source Kazakh speech dataset for TTS applications.
The KazakhTTS dataset contains over 93 hours of speech data (36 hours by female and 57 hours by male) consisting of around 42,000 recordings.
We released our dataset under the Creative Commons Attribution 4.0 International License, which permits both academic and commercial use.
We shared our experience by describing the dataset construction and TTS evaluation procedures, which might benefit other researchers planning to collect speech data for other low-resource languages.
Furthermore, the presented dataset can also aid to study and build (e.g., pretrain) TTS systems for other Turkic languages.
To demonstrate the use of our dataset, we built E2E-TTS models based on Tacotron 2 and Transformer architectures.
The subjective evaluation results suggest that the E2E-TTS models trained on KazakhTTS are suitable for practical use. 
We also shared the pretrained TTS models and the ESPnet training recipes for both speakers\textsuperscript{\ref{ft:github}}.

In future work, we plan to further extend our dataset by collecting more data from different domains, such as Wikipedia articles and books, and introduce new speakers.
We also plan to explore the optimal hyper-parameter settings for Kazakh E2E-TTS models, compare different TTS architectures, and conduct additional analysis.

\section{Acknowledgements}
We would like to thank our senior moderator Aigerim Boranbayeva for helping to train and monitor the transcribers. 
We also thank our technical editor Rustem Yeshpanov, PR manager Kuralay Baimenova, project cordinator Yerbol Absalyamov, and administration manager Gibrat Kurmanov for helping with other administrative and technical tasks.

\bibliographystyle{IEEEtran}

\bibliography{main}

\begin{thebibliography}{10}
\providecommand{\url}[1]{#1}
\csname url@samestyle\endcsname
\providecommand{\newblock}{\relax}
\providecommand{\bibinfo}[2]{#2}
\providecommand{\BIBentrySTDinterwordspacing}{\spaceskip=0pt\relax}
\providecommand{\BIBentryALTinterwordstretchfactor}{4}
\providecommand{\BIBentryALTinterwordspacing}{\spaceskip=\fontdimen2\font plus
\BIBentryALTinterwordstretchfactor\fontdimen3\font minus
  \fontdimen4\font\relax}
\providecommand{\BIBforeignlanguage}[2]{{%
\expandafter\ifx\csname l@#1\endcsname\relax
\typeout{** WARNING: IEEEtran.bst: No hyphenation pattern has been}%
\typeout{** loaded for the language `#1'. Using the pattern for}%
\typeout{** the default language instead.}%
\else
\language=\csname l@#1\endcsname
\fi
#2}}
\providecommand{\BIBdecl}{\relax}
\BIBdecl

\bibitem{taylor2009text}
P.~Taylor, \emph{Text-to-speech synthesis}.\hskip 1em plus 0.5em minus
  0.4em\relax Cambridge University Press, 2009.

\bibitem{dave2007kazakhstan}
B.~Dave, \emph{Kazakhstan-ethnicity, language and power}.\hskip 1em plus 0.5em
  minus 0.4em\relax Routledge, 2007.

\bibitem{DBLP:journals/corr/WangSSWWJYXCBLA17}
Y.~Wang, R.~J. Skerry{-}Ryan, D.~Stanton, Y.~Wu, R.~J. Weiss, N.~Jaitly,
  Z.~Yang, Y.~Xiao, Z.~Chen, S.~Bengio, Q.~V. Le, Y.~Agiomyrgiannakis,
  R.~Clark, and R.~A. Saurous, ``Tacotron: Towards end-to-end speech
  synthesis,'' in \emph{Proc. Annual Conference of the International Speech
  Communication Association (Interspeech)}.\hskip 1em plus 0.5em minus
  0.4em\relax {ISCA}, 2017, pp. 4006--4010.

\bibitem{DBLP:conf/iclr/SoteloMKSKCB17}
J.~Sotelo, S.~Mehri, K.~Kumar, J.~F. Santos, K.~Kastner, A.~C. Courville, and
  Y.~Bengio, ``{Char2Wav}: End-to-end speech synthesis,'' in \emph{Proc.
  International Conference on Learning Representations ({ICLR})}, 2017.

\bibitem{DBLP:conf/icml/ArikCCDGKLMNRSS17}
S.~{\"{O}}. Arik, M.~Chrzanowski, A.~Coates, G.~F. Diamos, A.~Gibiansky,
  Y.~Kang, X.~Li, J.~Miller, A.~Y. Ng, J.~Raiman, S.~Sengupta, and M.~Shoeybi,
  ``Deep {Voice}: Real-time neural text-to-speech,'' in \emph{Proc.
  International Conference on Machine Learning ({ICML})}, vol.~70.\hskip 1em
  plus 0.5em minus 0.4em\relax {PMLR}, 2017, pp. 195--204.

\bibitem{DBLP:conf/icml/Skerry-RyanBXWS18}
R.~J. Skerry{-}Ryan, E.~Battenberg, Y.~Xiao, Y.~Wang, D.~Stanton, J.~Shor,
  R.~J. Weiss, R.~Clark, and R.~A. Saurous, ``Towards end-to-end prosody
  transfer for expressive speech synthesis with tacotron,'' in \emph{Proc.
  International Conference on Machine Learning ({ICML})}, vol.~80.\hskip 1em
  plus 0.5em minus 0.4em\relax {PMLR}, 2018, pp. 4700--4709.

\bibitem{DBLP:conf/icassp/ShenPWSJYCZWRSA18}
J.~Shen, R.~Pang, R.~J. Weiss, M.~Schuster, N.~Jaitly, Z.~Yang, Z.~Chen,
  Y.~Zhang, Y.~Wang, R.~Ryan, R.~A. Saurous, Y.~Agiomyrgiannakis, and Y.~Wu,
  ``Natural {TTS} synthesis by conditioning {Wavenet} on {MEL} spectrogram
  predictions,'' in \emph{Proc. {IEEE} International Conference on Acoustics,
  Speech and Signal Processing ({ICASSP})}.\hskip 1em plus 0.5em minus
  0.4em\relax {IEEE}, 2018, pp. 4779--4783.

\bibitem{DBLP:conf/nips/RenRTQZZL19}
Y.~Ren, Y.~Ruan, X.~Tan, T.~Qin, S.~Zhao, Z.~Zhao, and T.~Liu, ``{FastSpeech}:
  Fast, robust and controllable text to speech,'' in \emph{Proc. Annual
  Conference on Neural Information Processing Systems (NeurIPS)}, 2019, pp.
  3165--3174.

\bibitem{DBLP:journals/corr/abs-1809-08895}
N.~Li, S.~Liu, Y.~Liu, S.~Zhao, and M.~Liu, ``Neural speech synthesis with
  transformer network,'' in \emph{Proc. {AAAI} Conference on Artificial
  Intelligence ({AAAI})}.\hskip 1em plus 0.5em minus 0.4em\relax {AAAI} Press,
  2019, pp. 6706--6713.

\bibitem{ljspeech17}
K.~Ito and L.~Johnson, ``The {LJ} speech dataset,''
  \url{https://keithito.com/LJ-Speech-Dataset/}, 2017.

\bibitem{DBLP:conf/interspeech/ZenDCZWJCW19}
H.~Zen, V.~Dang, R.~Clark, Y.~Zhang, R.~J. Weiss, Y.~Jia, Z.~Chen, and Y.~Wu,
  ``{LibriTTS}: A corpus derived from {LibriSpeech} for text-to-speech,'' in
  \emph{Proc. Annual Conference of the International Speech Communication
  Association ({Interspeech})}.\hskip 1em plus 0.5em minus 0.4em\relax {ISCA},
  2019, pp. 1526--1530.

\bibitem{DBLP:journals/corr/abs-2010-11567}
\BIBentryALTinterwordspacing
Y.~Shi, H.~Bu, X.~Xu, S.~Zhang, and M.~Li, ``{AISHELL-3:} {A} multi-speaker
  {Mandarin} {TTS} corpus and the baselines,'' \emph{CoRR}, vol.
  abs/2010.11567, 2020. [Online]. Available:
  \url{https://arxiv.org/abs/2010.11567}
\BIBentrySTDinterwordspacing

\bibitem{veaux2016superseded}
C.~Veaux, J.~Yamagishi, K.~MacDonald \emph{et~al.}, ``Superseded-{CSTR VCTK}
  corpus: English multi-speaker corpus for {CSTR} voice cloning toolkit,''
  2016.

\bibitem{DBLP:conf/interspeech/BlackT05}
A.~W. Black and K.~Tokuda, ``The {Blizzard Challenge} - 2005: Evaluating
  corpus-based speech synthesis on common datasets,'' in \emph{Proc. Eurospeech
  European Conference on Speech Communication and Technology
  ({Interspeech})}.\hskip 1em plus 0.5em minus 0.4em\relax {ISCA}, 2005, pp.
  77--80.

\bibitem{DBLP:conf/icml/RenTQZZL19}
Y.~Ren, X.~Tan, T.~Qin, S.~Zhao, Z.~Zhao, and T.~Liu, ``Almost unsupervised
  text to speech and automatic speech recognition,'' in \emph{Proc.
  International Conference on Machine Learning ({ICML})}, vol.~97.\hskip 1em
  plus 0.5em minus 0.4em\relax {PMLR}, 2019, pp. 5410--5419.

\bibitem{DBLP:conf/icassp/ChungWHZS19}
Y.~Chung, Y.~Wang, W.~Hsu, Y.~Zhang, and R.~J. Skerry{-}Ryan, ``Semi-supervised
  training for improving data efficiency in end-to-end speech synthesis,'' in
  \emph{Proc. {IEEE} International Conference on Acoustics, Speech and Signal
  Processing ({ICASSP})}.\hskip 1em plus 0.5em minus 0.4em\relax {IEEE}, 2019,
  pp. 6940--6944.

\bibitem{DBLP:conf/interspeech/ChenTYL19}
Y.~Chen, T.~Tu, C.~Yeh, and H.~Lee, ``End-to-end text-to-speech for
  low-resource languages by cross-lingual transfer learning,'' in \emph{Proc.
  Annual Conference of the International Speech Communication Association
  ({Interspeech})}.\hskip 1em plus 0.5em minus 0.4em\relax {ISCA}, 2019, pp.
  2075--2079.

\bibitem{khassanov2020crowdsourced}
\BIBentryALTinterwordspacing
Y.~Khassanov, S.~Mussakhojayeva, A.~Mirzakhmetov, A.~Adiyev, M.~Nurpeiissov,
  and H.~A. Varol, ``A crowdsourced open-source {Kazakh} speech corpus and
  initial speech recognition baseline,'' \emph{CoRR}, vol. abs/2009.10334,
  2020. [Online]. Available: \url{https://arxiv.org/abs/2009.10334}
\BIBentrySTDinterwordspacing

\bibitem{DBLP:conf/emnlp/MakhambetovMYMSS13}
O.~Makhambetov, A.~Makazhanov, Z.~Yessenbayev, B.~Matkarimov, I.~Sabyrgaliyev,
  and A.~Sharafudinov, ``Assembling the {Kazakh} language corpus,'' in
  \emph{Proc. Conference on Empirical Methods in Natural Language Processing
  ({EMNLP})}.\hskip 1em plus 0.5em minus 0.4em\relax {ACL}, 2013, pp.
  1022--1031.

\bibitem{DBLP:conf/aciids/MamyrbayevAZTG20}
O.~Mamyrbayev, K.~Alimhan, B.~Zhumazhanov, T.~Turdalykyzy, and F.~Gusmanova,
  ``End-to-end speech recognition in agglutinative languages,'' in \emph{Proc.
  Asian Conference on Intelligent Information and Database Systems ({ACIIDS})},
  vol. 12034, 2020, pp. 391--401.

\bibitem{boersma2001praat}
P.~Boersma, ``Praat, a system for doing phonetics by computer,'' \emph{Glot
  International}, vol.~5, no.~9, pp. 341--345, 2001.

\bibitem{hayashi2020espnet}
T.~Hayashi, R.~Yamamoto, K.~Inoue, T.~Yoshimura, S.~Watanabe, T.~Toda,
  K.~Takeda, Y.~Zhang, and X.~Tan, ``{ESPnet-TTS}: Unified, reproducible, and
  integratable open source end-to-end text-to-speech toolkit,'' in \emph{Proc.
  IEEE International Conference on Acoustics, Speech and Signal Processing
  ({ICASSP})}.\hskip 1em plus 0.5em minus 0.4em\relax IEEE, 2020, pp.
  7654--7658.

\bibitem{DBLP:conf/waspaa/PerraudinBS13}
N.~Perraudin, P.~Bal{\'{a}}zs, and P.~L. S{\o}ndergaard, ``A fast {Griffin-Lim}
  algorithm,'' in \emph{Proc. {IEEE} Workshop on Applications of Signal
  Processing to Audio and Acoustics ({WASPAA})}.\hskip 1em plus 0.5em minus
  0.4em\relax {IEEE}, 2013, pp. 1--4.

\bibitem{DBLP:conf/ssw/OordDZSVGKSK16}
A.~van~den Oord, S.~Dieleman, H.~Zen, K.~Simonyan, O.~Vinyals, A.~Graves,
  N.~Kalchbrenner, A.~W. Senior, and K.~Kavukcuoglu, ``{WaveNet}: A generative
  model for raw audio,'' in \emph{Proc. {ISCA} Speech Synthesis
  Workshop}.\hskip 1em plus 0.5em minus 0.4em\relax {ISCA}, 2016, p. 125.

\bibitem{DBLP:conf/icassp/YamamotoSK20}
R.~Yamamoto, E.~Song, and J.~Kim, ``Parallel {Wavegan}: {A} fast waveform
  generation model based on generative adversarial networks with
  multi-resolution spectrogram,'' in \emph{Proc. {IEEE} International
  Conference on Acoustics, Speech and Signal Processing ({ICASSP})}.\hskip 1em
  plus 0.5em minus 0.4em\relax {IEEE}, 2020, pp. 6199--6203.

\bibitem{DBLP:journals/corr/KingmaB14}
D.~P. Kingma and J.~Ba, ``Adam: {A} method for stochastic optimization,'' in
  \emph{Proc. International Conference on Learning Representations ({ICLR})},
  2015.

\bibitem{telegram}
\BIBentryALTinterwordspacing
{Telegram FZ LLC and Telegram Messenger Inc.}, ``Telegram.'' [Online].
  Available: \url{https://telegram.org}
\BIBentrySTDinterwordspacing

\bibitem{MTurk}
\BIBentryALTinterwordspacing
{Amazon.com, Inc.}, ``{Amazon Mechanical Turk (MTurk)}.'' [Online]. Available:
  \url{https://www.mturk.com}
\BIBentrySTDinterwordspacing

\end{thebibliography}


\end{document}